\documentclass[apj,iop,tighten]{emulateapj}
\usepackage{epsfig,aas_macros}
\usepackage{amsmath,amssymb,bm}
\usepackage{color}
\usepackage{aas_macros}
\usepackage[varg]{txfonts}

\begin{document}
\label{firstpage}

\title[Why are pulsar planets rare?]{Why are pulsar planets rare?}

\author{Rebecca G. Martin} 
\author{Mario Livio}
\author{Divya Palaniswamy}
\affil{Department of Physics and Astronomy,
  University of Nevada, Las Vegas, 4505 South Maryland Parkway, Las
  Vegas, NV 89154, USA }

\begin{abstract}
Pulsar timing observations have revealed planets around only a few
pulsars.  We suggest that the rarity of these planets is due mainly to
two effects. First, we show that the most likely formation mechanism
requires the destruction of a companion star. Only pulsars with a
suitable companion (with an extreme mass ratio) are able to form
planets.  Second, while a dead zone (a region of low turbulence) in
the disk is generally thought to be essential for planet formation, it
is most probably rare in disks around pulsars because of the
irradiation from the pulsar. The irradiation strongly heats the inner
parts of the disk pushing the inner boundary of the dead zone out.  We
suggest that the rarity of pulsar planets can be explained by the low
probability for these two requirements -- a very low--mass companion
and a dead zone -- to be satisfied.
\end{abstract}


\section{Introduction}
\label{intro}

\begin{figure}
\epsscale{1.0}
\begin{center}
\includegraphics[width=8.4cm]{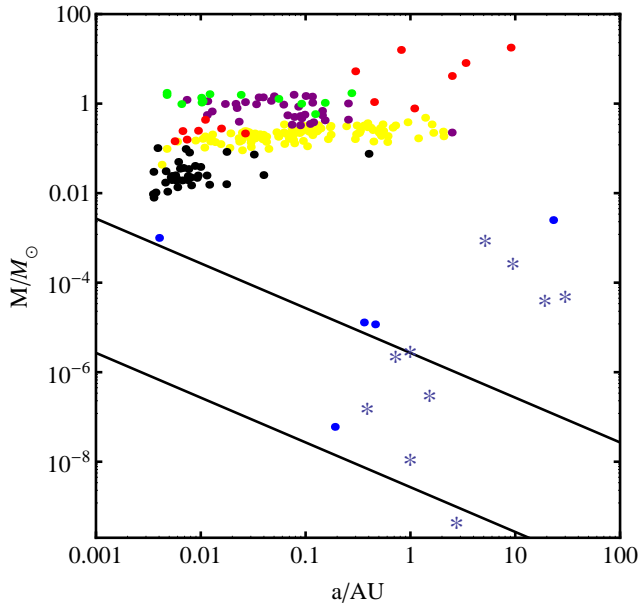}
\end{center}
\caption{Pulsar companion mass and semi--major axis. The mass is the
  median mass assuming an orbital inclination of $i=60^\circ$.  The
  assumed primary pulsar mass is $1.35\,\rm M_{\odot}$.  The
  companions are planets (blue points), main--sequence stars (red
  points), ultra low-mass (black points), neutron stars (green
  points), CO or ONeMg white dwarfs (purple points) or helium white
  dwarfs (yellow points). The blue stars show the 8 planets in the
  solar system, the Moon, and the asteroid Ceres. The solid lines show
  the detection limit for timing residuals of $1 \, \rm \mu s$ (lower
  line, applicable for millisecond pulsars) and $1\,\rm ms$ (upper
  line, applicable for normal slow pulsars) found with
  equation~(\ref{mass}). \label{Mass_semi}}
\end{figure}

There are currently five exoplanets in three planetary systems that
have been detected through pulsar timing to be orbiting pulsars,
rapidly rotating highly magnetized neutron stars
\citep{Lorimer2008}. The masses and semi--major axes of these planets
are shown in Table~\ref{tab} and plotted in the blue points in
Fig.~\ref{Mass_semi}. The first planets to be discovered (and
immediately confirmed) outside of our solar system were in fact found
around the pulsar PSR B1257+12 \citep{Wolszczan1992,Wolszczan1994,
  Wolszczan2012}. This pulsar has three very close--in planets. The
outer two planets are coplanar within $6^\circ$ and all three have low
eccentricity, implying a disk origin \citep{Konacki2003}. The outer
two planets are close to a 3:2 mean--motion resonance.  This suggests
that the planets did not form at their current location but instead
probably migrated in from farther out, possibly through a gas disk
\citep[e.g.][]{Terquem2007}.  A planet has also been observed around
the pulsar PSR J1719--1438. This planet has a mass similar to Jupiter
but a radius of less than about 40\% \citep{Bailes2011}. It is thought
to be an ultra--low mass white dwarf companion that has narrowly
avoided complete destruction \citep{VanHaaften2012}. Finally, a planet
has been detected around the pulsar PSR B1620-26
\citep{Backer1993,Sigurdsson2003}. This pulsar is a part of a binary
star system with a white dwarf and the planet is in a circumbinary
orbit. The most likely formation mechanism for this pulsar is that a
star and planet were captured by the pulsar, whose original companion
was ejected into space and lost \citep{Rasio1994,Ford2000pp}. Thus,
this planet probably did not form around the pulsar. All of the pulsar
planets found so far are around old millisecond pulsars (MSPs) that
are thought to have been spun up by accretion of matter from a
companion star \citep[e.g.][]{Alpar1982,Bhattacharya1991}.

\begin{table}
\caption{Mass and semi--major axis of planets observed around pulsars}
\centering
\begin{tabular}{lllcc}
\hline \hline
 Planet  & $M_{\rm planet}$ & $a$/AU \\ \hline
PSR B1257+12 A  & $0.02\,\rm M_\oplus$ & 0.19  \\
PSR B1257+12 B  & $4.3\,\rm M_\oplus$ & 0.36 \\
PSR B1257+12 C  & $3.9\,\rm M_\oplus$ & 0.46 \\
\hline
PSR J1719-1438 b  & $1\,\rm M_{\rm J}$ & 0.004 \\
\hline
PSR B1620--26 b  & $2.5\,\rm M_{\rm J}$ & 23  \\
\hline
\end{tabular}
\label{tab}
\end{table}

The precision of the pulsar timing allows detections of very low mass
bodies outside of the solar system \citep{Wolszczan1994,
  Wolszczan1997}.  The lower limit on the mass of an observable planet
is
\begin{equation}
M_{\rm planet} \, \sin(i) \approx 0.90 \, \left(\frac{\tau_{\rm
    pl}}{1\,\rm ms} \right)\, \left(\frac{a}{1\,\rm
  AU}\right)^{-1}\,\rm M_\oplus
\label{mass}
\end{equation}
\citep{Wolszczan1997}, where $\rm M_{\oplus}$ is the mass of the
Earth, $\tau_{\rm pl}$ is the measured timing residual amplitude, $a$
is the orbital semi--major axis and $i$ is the orbital inclination
\citep[e.g.][]{Thorsett1992,Blandford1993,Bailes1993,Wolszczan1997}. The
pulsar is assumed to have a mass $M_{\rm p}=1.35\,\rm M_\odot$.  We
show the detectability limits in Figure \ref{Mass_semi} for timing
residuals of $1\,\rm \mu s$ (applicable to MSPs) and $1\,\rm ms$
(applicable to normal pulsars) and compare to objects in the solar
system. Jovian planets are easily detectable around a normal slow
pulsar, whereas terrestrial planets, down to the size of large
asteroids, can be found in the timing residual of a MSP.

Despite the successes in finding planets around pulsars and the
precision offered, pulsar planets are rare
\citep[e.g.][]{Bailes1996,Bell1997pp}.  In particular, the survey of
\cite{Kerr2015} of 151 young pulsars revealed no planets.  Similarly,
the Australian Telescope National Facility (ATNF) pulsar
catalogue\footnote{http://www.atnf.csiro.au/people/pulsar/psrcat/}
currently contains $2536$ pulsars of which 436 are MSPs with spin
period $<10\,\rm ms$ \citep{ATNF2005}. Less than 1\% of observed MSPs
have a planetary mass companion.

In Section~2 we discuss different potential pulsar planet formation
models and find that the most likely place for a pulsar planet to form
is in a disk formed by the destruction of a binary companion.  In
Section~3 we discuss the probability for the formation of an extreme
mass ratio binary required for pulsar planet formation. In addition,
most planet formation scenarios require the presence of a ``dead
zone'' in the protoplanetary disk. This is a quiescent region where
the magneto--rotational instability (MRI) is unable to drive
turbulence and angular momentum transport
\citep[e.g.][]{Gammie1996,Gammie1998,Currie2007}. The dead zone is
thought to be a necessary component for planet formation for several
reasons. First, it provides a quiescent region where solids can settle
to the midplane \citep[e.g.][]{Youdin2002,
  Youdin2007,Zsom2011}. Second, it allows planetesimals to form
\citep[e.g.][]{Chambers2010,Bai2010,Youdin2011}. Finally, it slows the
rate of low--mass planet migration, preventing them from falling into
the star \citep[e.g.][]{Ward1997,Thommes2005,Matsumura2006}.  In
Section~4 we consider whether a dead zone can form in a disk around a
pulsar.  We discuss and summarize our results in Sections~5 and~6.

\section{Pulsar planet formation models}

There are several proposed mechanisms for planet formation around a
pulsar \citep[e.g.][]{Podsiadlowski1993,Podsiadlowski1995}. In this
section we discuss each in turn.

\subsection{Planets that survive the supernova?}

It is worth considering whether it is possible (at least in principle)
for planets to form in the usual way, in a protoplanetary disk around
a young (massive) star, and survive the pulsar formation process
\citep[e.g.][]{Bailes1991}. However, there is some doubt as to whether
planets can even form around a star massive enough to become a
pulsar. There is a strong decline in the probability of a star hosting
a giant planet for mass greater than $3\,\rm M_\odot$
\citep[e.g.][]{Kennedy2008}. Currently, the most massive star to have
a detected planet is $3\,\rm M_\odot$ \citep{Han2014}. This dearth of
planets around massive stars could be the result of several
factors. First, there is intense UV radiation from the star that can
rapidly destroy the disk. Second, the lifetime of the star is much
shorter. We should note, however, that selection effects may be
important since the planets (if they exist) would probably be
impossible to detect. Because the star is more massive and brighter
than a low mass star, neither the transit method nor the radial
velocity method is sensitive enough to detect a planet.

Supposing that such planets can still form, then there are several
conditions that must be met in order for them to survive the death of
the star. The orbital radii of the planets must be larger than about
$4\,\rm AU$ in order to survive being engulfed during the red--giant
phase.  In the supernova explosion, half of the mass of the system is
ejected. In order for planets to survive, the explosion must be
asymmetric. The asymmetry results in a velocity kick to the newly
formed pulsar that must be in a similar direction to the motion of the
planet at the time of the supernova \citep{Blaauw1961,
  Bhattacharya1991}. Planets that survive would be expected to be in
eccentric orbits at orbital radii of at least a few AU
\citep[e.g.][]{Thorsett1993}.

This mechanism is unlikely to form a system with more than one planet
(if any). For three planets to survive this process is almost
certainly impossible and so the planets around PSR B1257+12 formed
after the pulsar. Furthermore, the two other known pulsar planets most
likely did not form in this way since PSR B1620--26 is a circumbinary
planet and PSR J1719--1438 is extremely close to its host star. None
of the known pulsar planets formed in this way and we suggest that the
probability for planets to form in this scenario is negligible.

\subsection{Supernova fallback disk?}
\label{22}

When a pulsar forms in a supernova, most of the material of the star
is ejected. However, some material can fall back towards the newly
formed pulsar if it does not attain the escape velocity or if it is
pushed back during the early hydrodynamical mixing phase
\citep[e.g.][]{Colgate1971, Chevalier1989,Bailes1991}. Estimates of
the total mass that can form the fall back disk are in the range
$0.001-0.1\,\rm M_\odot$ \citep[e.g.][]{Lin1991, Menou2001}. This is
similar to the masses observed in a protoplanetary disks around young
stars that lie in the range $0.001-0.1\,\rm M_\star$
\citep[e.g.][]{Williams2011}, where $M_\star$ is the mass of the host
star. However, the angular momentum of the fallback disk may be very
different. Recently, \cite{Perna2014} considered the formation of a
supernova fallback disk around neutron stars (and black holes) with
numerical simulations of supernova explosions. They found that if
magnetic torques play a role in angular momentum transport, then
supernova fallback disks do not form around neutron stars. Any
material that falls back has too little angular momentum to orbit the
neutron star even once. However, formation can occur when the magnetic
torques are negligible. This requires the energy of the explosion to
be finely tuned so that the fall back mass would on one hand not cause
the neutron star to collapse to a black hole, but on the other would
be sufficient to form a disk. \cite{Perna2014} found a typical
fallback mass of $0.08\,\rm M_\odot$ with the maximum specific angular
momentum of the material $\lesssim 10^{17}\,\rm cm^2\,s^{-1}$.

We can calculate the circularization radius for such a disk. The
radius of a ring of material with a given specific angular momentum,
$j$, is
\begin{equation}
R_{\rm circ}=\frac{j^2}{GM_{\rm p}},
\end{equation}
where $M_{\rm p}$ is the mass of the pulsar. For typical parameters this is
\begin{equation}
R_{\rm circ}=5.4\times 10^7\left(\frac{j}{10^{17}\,\rm erg\, s}\right)^2
\left(\frac{M_{\rm p}}{1.4\,\rm M_\odot}\right)^{-1}\,\rm cm.
\end{equation}
The upper limit to the total angular momentum of such a disk is
  estimated to be 
\begin{equation}
J=2\times 10^{49}\left(\frac{M_{\rm d}}{0.1\,\rm
  M_\odot}\right)\left(\frac{j}{10^{17}\,\rm erg\, s}\right)\,\rm erg
\, s,
\label{sf}
\end{equation}
where $M_{\rm d}$ is the mass of the disk.  This is in line with
estimates by \cite{Phinney1993} and \cite{Menou2001}.  A simple
estimate for the lifetime of such a disk is the viscous timescale
\begin{equation}
\tau_\nu=\frac{R^2}{\nu},
\label{tauvisc}
\end{equation}
where the viscosity is
\begin{equation}
\nu=\alpha \left(\frac{H}{R}\right)^2 R^2 \Omega,
\label{nu}
\end{equation}
$H/R$ is the disk aspect ratio, $\Omega=\sqrt{GM_{\rm p}/R^3}$ is the
Keplerian angular frequency and $\alpha$ is the \cite{SS1973}
viscosity parameter.  For typical values at the circularization radius
we find
\begin{align}
\tau_\nu(R_{\rm circ})=& \,\, 2.8\times 10^4
\left(\frac{\alpha}{0.01}\right)^{-1}
\left(\frac{H/R}{0.01}\right)^{-2} \left(\frac{M_{\rm p}}{1.4\,\rm
  M_\odot}\right)^{-2}\cr
& \,\, \left(\frac{j}{10^{17}\,\rm erg\, s}\right)^{3}\,\rm s.
\end{align}
The viscous disk spreads both inwards and outwards leading to a longer
lifetime for larger radii. However, even an increase in this timescale
by several orders of magnitude would not be sufficient for a
reasonable timescale for planet formation. Even taking into
account uncertainties in some of the physical parameters, if a
supernova fallback disk does form, its lifetime is far too short for
planet formation to occur.

Observations aimed at finding fallback disks around pulsars have not
been successful \citep[e.g.][]{Wang2007}. This, combined with the lack
of observed planets around young pulsars \citep{Kerr2015} suggests
that planets do not form in a supernova fallback disk. The observed
planets more likely formed during a later accretion phase that led to
the spin up of the pulsar.

\subsection{Destruction of a companion star}
\label{dis}

This model involves a close binary composed of a MSP (or its
progenitor) and a low--mass main--sequence star or compact companion
\citep[e.g.][]{Podsiadlowski1991,Stevens1992}. The companion is losing
mass through evaporation because it is being irradiated by the pulsar.
If the evaporation timescale is shorter than the Kelvin--Helmholtz
timescale the secondary responds adiabatically to the mass loss.  The
companion star fills its Roche lobe and is dynamically disrupted if
its radius increases faster than the Roche--lobe radius
\citep[e.g.][]{Benz1990}. This is most likely to happen when the
companion is fully convective or degenerate (e.g. for a low mass star
the radius of the star is proportional to $M^{-1/3}$, where $M$ is the
mass of the star). The star is dynamically disrupted and forms a
massive disk around the primary with a mass of around $0.1\,\rm
M_\odot$.

In the specific case of the merger of two white dwarfs, the orbit
shrinks due to gravitational radiation until the less massive white
dwarf fills its Roche lobe. However, in this case, there is some
uncertainty whether the merger will lead to a type~Ia supernova rather
than to the formation of a pulsar \citep[e.g.][]{Livio2000}. It
is most likely that a pulsar must be a component of the binary before
the destruction of the companion.

During the dissolution of a companion star, the material forms a disk
at a radius comparable to the initial separation of the binary, but
then spreads out through viscous effects. Note that the disk formed is
much larger than that expected from a supernova fallback disk. 
The total angular momentum of the disk is
\begin{equation}
J=1.1\times 10^{52}\left(\frac{M_{\rm d}}{0.1\,\rm M_\odot}\right)
\left(\frac{M_{\rm p}}{1.4\,\rm M_\odot}\right)^\frac{1}{2}
\left(\frac{a}{1\,\rm AU}\right)^\frac{1}{2}\,\rm erg\, s
\end{equation}
This is several orders of magnitude higher than the maximum estimate
for the angular momentum of the supernova fallback disk given in
equation~(\ref{sf}).  With the viscous timescale given by
equation~(\ref{tauvisc}) and the viscosity in equation~(\ref{nu}), at
a radius of $R=1\,\rm AU$, we find the viscous timescale to be
\begin{align}
\tau_\nu=1.34\times 10^5 
\left(\frac{\alpha}{0.01}\right)^{-1}
\left(\frac{H/R}{0.01}\right)^{-2}
\left(\frac{M_{\rm p}}{1.4\,\rm M_\odot}\right)^{-\frac{1}{2}}
\left(\frac{R}{1\,\rm AU}\right)^{\frac{3}{2}}\,\rm yr.
\end{align}
Since some material spreads outwards from its initial radius, the disk
lifetime is longer than this. For example, material that reaches
$R=5\,\rm AU$ has a viscous timescale of over $1\,\rm Myr$. This is
similar to the disk lifetime for a protoplanetary disk around a young
star \citep[e.g.][]{Williams2011}. We therefore suggest that the
destruction of a companion star is the most likely scenario for
forming the planetary system around PSR B1257+12.  The accretion rate
is typically $\dot M=M_{\rm d}/\tau_\nu$ which is
\begin{align}
\dot M=&\,\, 6.64\times 10^{-8}
\left(\frac{\alpha}{0.01}\right)
\left(\frac{H/R}{0.01}\right)^{2}
\left(\frac{M_{\rm d}}{0.1\,\rm M_\odot}\right)\cr
& \,\, \times
\left(\frac{M_{\rm p}}{1.4\,\rm M_\odot}\right)^\frac{1}{2}
\left(\frac{R}{5\,\rm AU}\right)^{-\frac{3}{2}}
\,\rm M_\odot \, yr^{-1}.
\end{align}
This is similar to typical accretion rates observed around young
T~Tauri stars \citep[e.g.][]{Armitage2003,Calvet2004}.  Over time, as
the mass of the disk decreases, the accretion rate also decreases. For
a fully viscous disk, once the material has spread out into the inner
regions, the disk can be modeled to be in a quasi--steady--state. In
Section~\ref{pd} we investigate the properties of this model further.

\subsection{Evaporation of a companion}

A companion star may be evaporated by the radiation from a pulsar
\citep{Fruchter1990,Bailes1991,Krolik1991,Rasio1992,Tavani1992}.  The
companion has to be almost, but not completely, evaporated by the
pulsar leaving a remnant that is planet sized. This is thought to be
the explanation for the planet in PSR J1719--1438. This mechanism is
only capable of forming a single planet around a pulsar and so this
scenario is unlikely to have been responsible for the planetary system
around PSR B1257+12. The fine tuning required for this mechanism for
planet formation means that this is probably a rare event. If the
companion is completely evaporated and the planets form from that
debris, then the formation is somewhat similar to that described in
the previous Section~\ref{dis}. However, the disk formed in this
manner may have very little mass and so planet formation may be
unlikely. 

In summary, the most likely mechanisms for pulsar planet formation
require a low mass companion to the pulsar. In the next Section we
examine the parameters required to form an extreme mass ratio binary
that is potentially capable of pulsar planet formation.

\section{Binary formation}

As we have shown in the previous section, planet formation around a
pulsar requires a low mass binary companion. A massive star that ends
its life as a neutron star has a mass in the approximate range
$9-25\,\rm M_\odot$ \citep{Heger2003}.  The binary fraction of high
mass stars may be quite high, up to 100\%
\citep[e.g.][]{Kratter2006,Mason2009}.  However, for pulsar planet
formation, the binary companion must have a mass such that when it is
accreted on to the pulsar it does not cause the latter to collapse to
form a black hole. If the companion is a fully convective low mass
main--sequence star, its mass is in the approximate range $0.1{\,\rm
  M_\odot}\lesssim M_2\lesssim 0.3\,\rm M_\odot$. Alternatively, the
companion may be a very low mass white dwarf, and the progenitor
star would have to have a mass less than a solar
mass. Observationally, the mass ratio distribution of binaries is
dependent on the mass of the primary star
\citep[e.g.][]{Bastian2010}. The binary mass ratio distribution for
high mass primary stars is not well constrained for low mass ratios,
$q\lesssim 0.1$. However, for $0.1<q<1$ the mass ratio distribution is
relatively flat \citep[e.g.][]{Duchene2013}. Assuming that this
extends to lower mass objects the probability of a binary forming with
a mass ratio required to form pulsar planets $q=M_2/M_1<0.1$ is
$\lesssim 10$\%.

When the pulsar formed in a supernova explosion, an asymmetry in the
explosion leads to a kick on the newly formed neutron star
\citep[e.g.][]{Shklovskii1970}. If the supernova kick is too strong,
the system does not remain bound
\citep[e.g.][]{Brandt1995,Martinetal2009b,Martinetal2010}. The smaller
the binary companion, the more likely the system is to become unbound.
For a binary system that can form pulsar planets, we estimate that the
probability of a bound orbit is $\lesssim 10$\% \citep[see Figure 2 in
][]{Brandt1995}.

In conclusion, we find that pulsar planet formation requires both an
extreme mass ratio binary and for this binary to survive the supernova
explosion that forms the pulsar. The probability of an extreme mass
ratio binary forming and surviving the supernova is very small.

\section{Pulsar Disk Models}
\label{pd}

The main difference between a protoplanetary disk around a young star
and a disk around a pulsar is that the pulsar strongly irradiates the
disk and provides an additional source of heating.  We consider
whether a dead zone forms in pulsar disks by considering the surface
density and temperature structure of a disk around a pulsar.

The MRI drives turbulence within an accretion disk that transports
angular momentum outwards allowing material to spiral inwards
\citep[e.g.][]{BH1991}. The MRI operates when the disk is sufficiently
ionized, which requires a temperature greater than the critical value,
$T>T_{\rm crit}$. The value of the critical temperature is thought to
be in the range $800-1400\,\rm K$
\citep[e.g.][]{Umebayashi1988,Zhuetal2010a}.  The disk temperature
decreases with radius. For temperatures lower than the critical value,
the disk is only fully MRI active if the surface density is
sufficiently small, $\Sigma<\Sigma_{\rm crit}$. If the surface density
is larger than the critical value, a dead zone forms at the midplane
and material only flows through the surface layers since these may be
ionized by external sources. Cosmic rays ionize about $200\,\rm g\,
cm^{-2}$ \citep[e.g.][]{Gammie1996,Fromangetal2002}. However, an MHD
jet or a disk wind can sweep away the cosmic rays
\citep[e.g.][]{Skilling1976,Cesarsky1978}. Low--energy cosmic rays
ionize more than higher energy cosmic rays \citep[e.g.][]{Lepp1992}
and these may be excluded from the disk by magnetic scattering. X-rays
can only penetrate a much smaller surface density of around $0.1\,\rm
g\, cm^{-2}$ \citep[e.g.][]{Matsumura2003,Glassgoldetal2004}.  Since
there is some uncertainty as to the critical value for the surface
density, we regard $\Sigma_{\rm crit}$ as a free parameter and
consider different values in this work \citep[see
  also][]{Armitage2001,Zhuetal2009,Martinetal2012a,Martinetal2012b}.

A disk with a dead zone is not in a steady state, as material builds
up there. The dead zone acts like a plug in the accretion flow and
material can only flow through the surface layers with surface
density, $\Sigma_{\rm crit}$. This can result in a massive disk that
is unstable to gravitational instability if the \cite{Toomre1964}
parameter becomes sufficiently small. This leads to gravitational
turbulence that heats the disk and can potentially trigger the MRI in
the dead zone. This causes an accretion outburst
\citep[e.g.][]{Armitage2001,Zhuetal2010b,
  MartinandLubow2011,MartinandLubow2013prop,MartinandLubow2013dza}. Around
a young star, an accretion outburst accretes around $0.1\,\rm
M_\odot$. Since this is the total mass of the disk around the pulsar
and this material spreads out, outbursts are unlikely around
pulsars. Thus we can consider just the location of a dead zone in a
disk around a pulsar.  We consider a quasi--steady--state disk model
which assumes that the disk is fully turbulent for each accretion
rate, but we determine when and where a dead zone exists.

\subsection{Inner edge of the dead zone}

If a dead zone forms, the inner edge, closest to the star, is
determined by where the temperature of the disk drops below the
critical temperature, $T<T_{\rm crit}$. In this section, we assume
that in the inner parts of the disk, the temperature is dominated by
the irradiation from the pulsar. The heating due to the irradiation is
\begin{equation}
Q_{\rm irr}=\sigma T_{\rm irr}^4=\frac{L}{4\pi R^2}(1-\beta)\cos \phi,
\label{qirr}
\end{equation}
\citep[e.g.][]{Franketal2002} where the albedo is $\beta=0.5$ and
\begin{equation}
\cos \phi=\frac{dH}{dR}-\frac{H}{R}. 
\label{cosphi}
\end{equation}
The accretion luminosity is
\begin{equation}
L_{\rm acc}=\frac{GM_{\rm p}\dot M}{R_{\rm p}},
\label{lacc}
\end{equation}
where $\dot M$ is the accretion rate on to the pulsar. For typical values this is
\begin{equation}
L_{\rm acc}=3.1\times 10^4 
\left(\frac{\dot M}{10^{-8}\,\rm M_\odot\, yr^{-1}}\right)
\left(\frac{M_{\rm p}}{1.4\,\rm M_\odot}\right)
\left(\frac{R_{\rm p}}{10\,\rm km}\right)^{-1}\,\rm L_{\odot}.
\end{equation}
However, this is limited by the Eddington luminosity,
\begin{equation}
L_{\rm EDD}=4.5\times 10^4 \left(\frac{M_{\rm p}}{1.4\,\rm M_\odot}\right)\,\rm L_\odot.
\label{ledd}
\end{equation}
Therefore, the luminosity in equation~(\ref{qirr}) is
\begin{equation}
L=\min (L_{\rm acc},L_{\rm EDD}).
\end{equation}
We find the critical accretion rate above which the luminosity is
limited by the Eddington Luminosity by solving $L_{\rm acc}=L_{\rm
  EDD}$ to be
\begin{equation}
\dot M_{\rm crit}=1.45\times 10^{-8}\left(\frac{R_{\rm p}}{10\,\rm km}\right)   \,\rm M_\odot\, yr^{-1}.
\end{equation}

\begin{figure}
\begin{center}
\includegraphics[width=8.4cm]{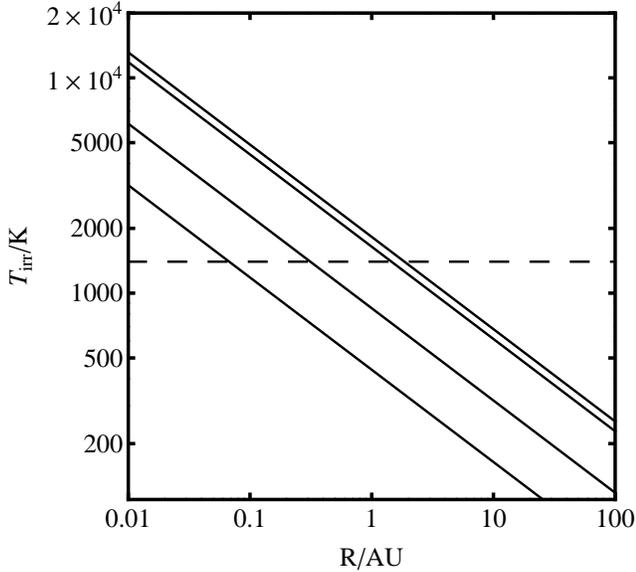}
\end{center}
\caption{Irradiation temperature as a function of radius from the
  pulsar for a steady--state disk for four different accretion
  rates. The upper solid line is for accretion rates of $\dot M>\dot
  M_{\rm crit}=1.45\times 10^{-8}\,\rm M_\odot\, yr^{-1}$ that are
  above the Eddington luminosity and the temperature is determined
  with equation~(\ref{tedd}). The irradiation temperature for the
  lower accretion rates is determined with equation~(\ref{t}). We show
  $\dot M=10^{-8}\,\rm M_\odot\, yr^{-1}$ (second solid line),
  $\dot M=10^{-9}\,\rm M_\odot\, yr^{-1}$ (third solid line) and $\dot
  M=10^{-10}\,\rm M_\odot\, yr^{-1}$ (lower solid line). The dashed
  line shows the critical temperature required for the MRI to operate
  of $T_{\rm crit}=1400\,\rm K$. The radius at which each the solid
  line crosses the dashed line is the inner edge of a dead zone, if it
  forms, $R_{\rm d,in}$.}
\label{tirr} 
\end{figure}

We assume that the disk aspect ratio can be written as a power law in
radius $H \propto R^n$, This simplifies equation~(\ref{cosphi}) to read
\begin{equation}
\cos \phi= \frac{H}{R}(n-1).
\end{equation}
Assuming hydrostatic equilibrium and that the disk temperature is
dominated by the irradiation, we have
\begin{equation}
H=\frac{c_{\rm s}}{\Omega}=\frac{1}{\Omega}\sqrt{\frac{{\cal R}T_{\rm irr}}{\mu}},
\label{h}
\end{equation}
where ${\cal R}=8.31\times 10^7\,\rm erg\,K^{-1}\, mol^{-1}$ is the
gas constant and $\mu=2.3$ is the mean molecular weight. We take
equation~(\ref{qirr}) (with $L=L_{\rm acc}$ defined in
equation~(\ref{lacc})) and combine it with equation~(\ref{h}) to write
the irradiation temperature for $L_{\rm acc}<L_{\rm EDD}$ as
\begin{align}
T_{\rm irr}=&\,\,1643
\left(\frac{\dot M}{10^{-8}\,\rm M_\odot\, yr^{-1}}\right)^\frac{2}{7}
\left(\frac{M_{\rm p}}{1.4\,\rm M_\odot}\right)^\frac{1}{7}\cr
& \,\, \times
\left(\frac{R_{\rm p}}{10\,\rm km}\right)^{-\frac{2}{7}}
\left(\frac{R}{1\,\rm AU}\right)^{-\frac{3}{7}}
\,\rm K.
\label{t}
\end{align}
Since we find $T_{\rm irr}\propto R^{-3/7}$, this gives $H=c_{\rm
  s}/\Omega \propto T_{\rm irr}^{1/2}R^{3/2} \propto R^{9/7}$.  For a
disk that is dominated by irradiation, $n=9/7$ and we use this value
for the rest of the work.  Similarly, for higher accretion rates, with
$L_{\rm acc}>L_{\rm EDD}$, we combine equation~(\ref{qirr}) (with
$L=L_{\rm EDD}$ in equation~(\ref{ledd})) with equation~(\ref{h}) and
we find
\begin{align}
T_{\rm irr, EDD}=&\,\,1825
\left(\frac{M_{\rm p}}{1.4\,\rm M_\odot}\right)^\frac{1}{7}
\left(\frac{R}{1\,\rm AU}\right)^{-\frac{3}{7}}
\,\rm K.
\label{tedd}
\end{align}
In Fig.~\ref{tirr} we show the disk temperature as a function of
radius for various accretion rates. We also show the critical
temperature required for the MRI to operate. A dead zone forms when
the temperature of the disk drops below this. We solve $T_{\rm
  irr}=T_{\rm crit}$ to find the critical radius outside of which the
temperature drops below that required for the MRI, 
\begin{align}
R_{\rm d,in}=& \,\,1.45
\left(\frac{\dot M}{10^{-8}\,\rm M_\odot\, yr^{-1}}\right)^\frac{2}{3}
\left(\frac{R_{\rm p}}{10\,\rm km}\right)^{-\frac{2}{3}}\cr
&\,\, \times
\left(\frac{M_{\rm p}}{1.4\,\rm M_\odot}\right)^\frac{1}{3}
 \left(\frac{T_{\rm c}}{1400\,\rm K}\right)^{-\frac{7}{3}}
\,\rm AU.
\label{up}
\end{align}
If, at this radius, the surface density is sufficiently high, then a
dead zone forms outside of this radius. Above the Eddington limit, we
solve $T_{\rm irr, EDD}=T_{\rm crit}$ to find the critical inner dead
zone radius 
\begin{align}
R_{\rm d,in,EDD}=& \,\,1.85
\left(\frac{M_{\rm p}}{1.4\,\rm M_\odot}\right)^\frac{1}{3}
 \left(\frac{T_{\rm c}}{1400\,\rm K}\right)^{-\frac{7}{3}}
\,\rm AU.
\label{down}
\end{align}
The lower line in Fig.~\ref{rcrit} shows the inner edge of the dead
zone as a function of accretion rate. This is a combination of
equation~(\ref{up}) for $\dot M<\dot M_{\rm crit}$ and
equation~(\ref{down}) for $\dot M>\dot M_{\rm crit}$. Since the inner
dead zone radius does not depend upon the critical surface density it
is the same in both plots.  In the next Section we consider the radial
location of the outer edge of the dead zone.

\subsection{Outer edge of the dead zone}

\begin{figure}
\begin{center}
\includegraphics[width=8.4cm]{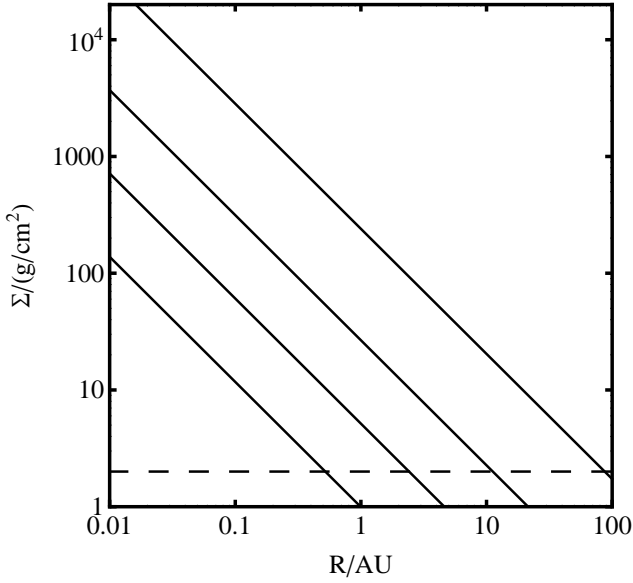}
\end{center}
\caption{Surface density as a function of radius from the pulsar for a
  steady--state disk for four different accretion rates. The upper
  solid line is above the Eddington limit and is determined with
  equation~(\ref{sigEDD}) and shows $\dot M=10^{-7}\,\rm M_\odot\,
  yr^{-1}$. The other solid lines show accretion rates below the
  Eddington limit that are determined with equation~(\ref{sig}), $\dot
  M=10^{-8}\,\rm M_\odot\, yr^{-1}$ (second solid line), $\dot
  M=10^{-9}\,\rm M_\odot\, yr^{-1}$ (third solid line) and $\dot
  M=10^{-10}\,\rm M_\odot\, yr^{-1}$ (lower solid line). The dashed
  line shows the critical surface density that is ionized by external
  sources, of $\Sigma_{\rm crit}=2\,\rm g\,cm^{-2}$. The radius at
  which the solid line crosses the dashed line is the outer edge of a
  dead zone, if it forms, $R_{\rm d,out}$.}
\label{sd}
\end{figure}

\begin{figure*}
\begin{center}
\includegraphics[width=8.4cm]{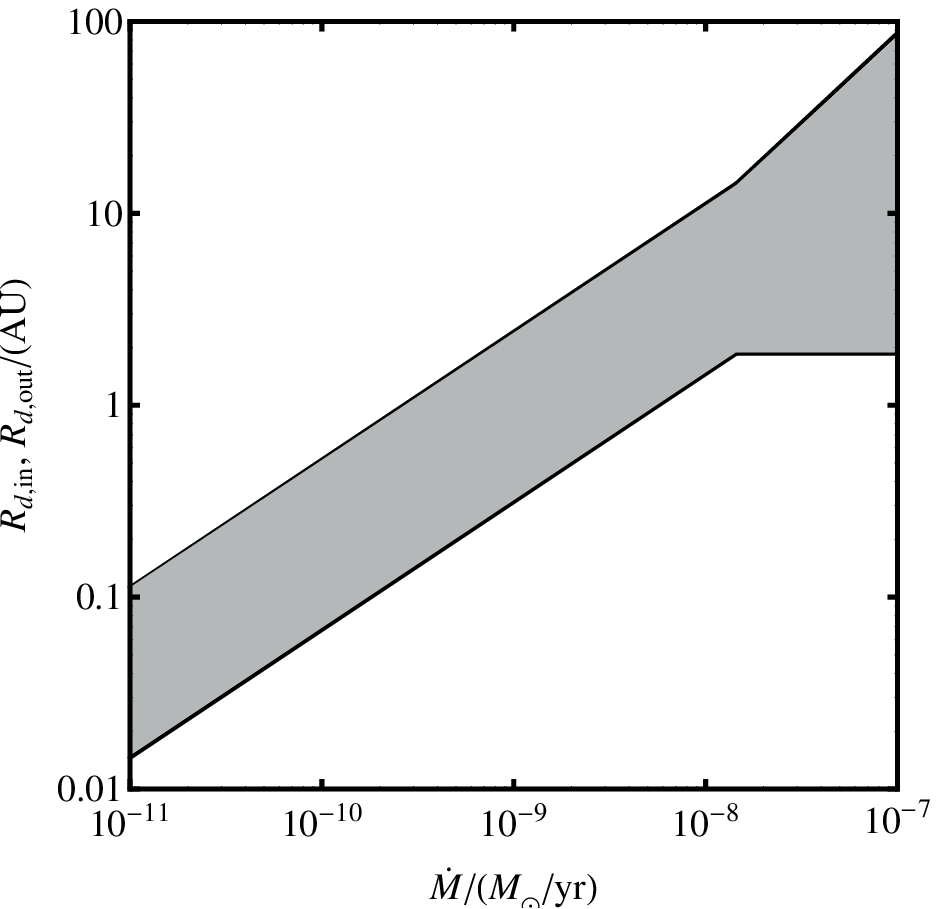}
\includegraphics[width=8.4cm]{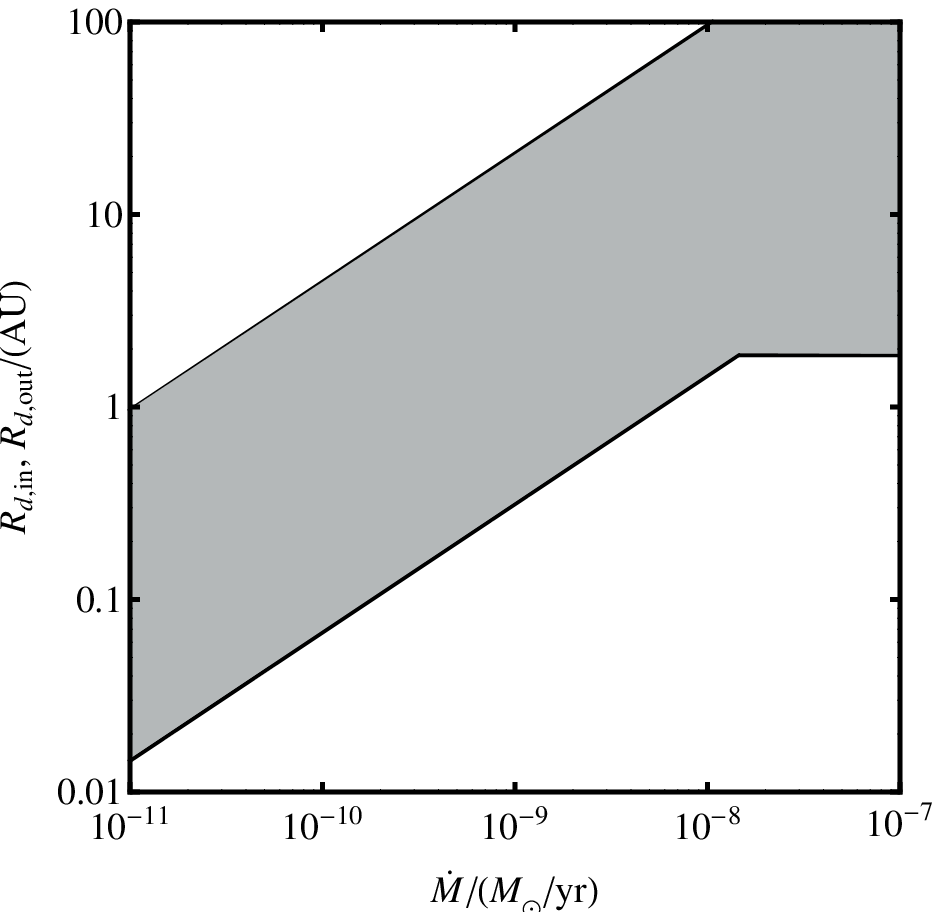}
\end{center}
\caption{Inner edge of the dead zone, $R_{\rm d,in}$ (defined where
  $T_{\rm irr}=T_{\rm crit}$) and the outer edge of the dead zone,
  $R_{\rm d,out}$, (defined where $\Sigma=\Sigma_{\rm crit}$). The
  dead zone exists in the shaded region. We take $T_{\rm
    crit}=1400\,\rm K$ and $\Sigma_{\rm crit}=2\,\rm g\, cm^{-2}$
  (left) and $\Sigma_{\rm crit}=0.2\,\rm g\, cm^{-2}$ (right).}
\label{rcrit}
\end{figure*}

If a dead zone is present, its outer edge is determined by the
location where the surface density drops below the critical,
$\Sigma<\Sigma_{\rm crit}$. The viscosity of the disk is
\begin{equation}
\nu=\alpha \frac{c_{\rm s}^2}{ \Omega}.
\end{equation}
For a steady state disk the surface density is given by
\begin{equation}
\Sigma=\frac{\dot M}{3\pi \nu}
\end{equation}
\citep{Pringle1981}.  For our typical parameters, for an accretion
rate that is not Eddington limited, this is
\begin{align}
\Sigma=&\,\,26.7
\left(\frac{\dot M}{10^{-8}\,\rm M_\odot\, yr^{-1}}\right)^\frac{5}{7}
\left(\frac{R_{\rm p}}{10\,\rm km}\right)^{\frac{2}{7}}
\left(\frac{\alpha}{0.01}\right)^{-1}
\left(\frac{M_{\rm p}}{1.4\,\rm M_\odot }\right)^{\frac{5}{14}} \cr
&\,\,\times
\left( \frac{R}{1\,\rm AU}\right)^{-\frac{15}{14}}
\,\rm g\, cm^{-2}.
\label{sig}
\end{align}
Similarly, for the case in which the luminosity is Eddington limited,
we can find the surface density
\begin{align}
\Sigma_{\rm EDD}=&\,\,24.0
\left(\frac{\dot M}{10^{-8}\,\rm M_\odot\, yr^{-1}}\right)
\left(\frac{\alpha}{0.01}\right)^{-1}
\left(\frac{M_{\rm p}}{1.4\,\rm M_\odot }\right)^{\frac{5}{14}} \cr
&\,\,\times
\left( \frac{R}{1\,\rm AU}\right)^{-\frac{15}{14}}
\,\rm g\, cm^{-2}.
\label{sigEDD}
\end{align}
In Fig.~\ref{sd} we plot the surface density as a function of radius
for various accretion rates. 

We can find the radius of the outer edge of the dead zone by solving
$\Sigma=\Sigma_{\rm crit}$. Below the Eddington accretion rate we find
\begin{align}
R_{\rm d,out} =& \,\,1.3
\left(\frac{\dot M}{10^{-8}\,\rm M_\odot\, yr^{-1}}\right)^\frac{2}{3}
\left(\frac{\alpha}{0.01}\right)^{-\frac{14}{15}}
\left(\frac{R_{\rm p}}{10\,\rm km}\right)^\frac{4}{15}\cr
&\,\, \times
\left(\frac{M_{\rm p}}{1.4\,\rm M_\odot }\right)^{\frac{1}{3}}
\left(\frac{\Sigma_{\rm crit}}{20\,\rm g\, cm^{-2}}\right)^{-\frac{14}{15}}
\,\rm AU
\end{align}
and the corresponding critical radius above the Eddington accretion
rate is
\begin{align}
R_{\rm d,out,EDD} =& \,\,1.18
\left(\frac{\dot M}{10^{-8}\,\rm M_\odot\, yr^{-1}}\right)^\frac{14}{15}
\left(\frac{\alpha}{0.01}\right)^{-\frac{14}{15}}\cr
&\,\, \times
\left(\frac{M_{\rm p}}{1.4\,\rm M_\odot }\right)^{\frac{1}{3}}
\left(\frac{\Sigma_{\rm crit}}{20\,\rm g\, cm^{-2}}\right)^{-\frac{14}{15}}
\,\rm AU.
\end{align}
In Fig.~\ref{rcrit} the upper lines show the outer edge of the
  dead zone that depends on both the accretion rate and the critical
  active layer surface density.  In the next Section, we determine
whether a dead zone can exist since if $R_{\rm d,in}>R_{\rm d,out}$,
then there is no region of parameter space for a dead zone and the
disk is fully turbulent.

\subsection{Existence of a dead zone}

The condition that must be satisfied for a dead zone to exist is that
the surface density at the radius of the inner dead zone boundary must
be larger than the critical surface density that is MRI active,
\begin{equation}
\Sigma (R_{\rm d,in})>\Sigma_{\rm crit}.
\end{equation}  
Since the surface density in the disk increases with radius, the
larger $\Sigma_{\rm crit}$ the less likely it is that a dead zone may
form. This condition is equivalent to
\begin{equation}
R_{\rm d,in}<R_{\rm d,out}.
\end{equation}
We consider here disk parameters for which a dead zone exists.

In Fig.~(\ref{rcrit}) we show the inner and outer dead zone radii for
an example with $T_{\rm crit}=1400\,\rm K$ and $\Sigma_{\rm
  crit}=2\,\rm g\, cm^{-2}$ (left) and $\Sigma_{\rm crit}=0.2\,\rm g\,
cm^{-2}$ (right). The shaded regions show the accretion rates for
which a dead zone exists and the range of radii.  We find that for
$\Sigma_{\rm crit}\gtrsim 10\,\rm g\, cm^{-2}$ a dead zone does not
exist for typical accretion rates.  However, for a smaller critical
surface density, $\Sigma_{\rm crit}\lesssim 10\,\rm g\, cm^{-2}$,
there is a dead zone for all accretion rates. We suggest that the
formation of planets around a pulsar requires a very small critical
surface density. Such a low critical surface density may be difficult
to achieve because cosmic rays are expected to penetrate $\Sigma_{\rm
  crit}=200\,\rm g\, cm^{-2}$. There are several possibilities for
lowering the active layer surface density. First, as described in
Section~\ref{pd}, cosmic rays may be swept away allowing a large dead
zone to form. Second, the presence of dust or polycyclic hydrocarbons
can suppress the MRI. Magnetohydrodynamic simulations that include
these effects find small critical surface densities
\citep[e.g.][]{Bai2011,Perezbecker2011}. Third, the inclusion of
non--ideal MHD effects may decrease the amount of turbulence
\citep[e.g.][]{Simon2013}. There remains much uncertainty in the value
of the active layer surface density.

We note that our model represents a steady state disk but in a real
disk around the pulsar there will initially be a spreading phase
before the steady state is reached. If the disk is made by the
destruction of a binary companion, then, initially the material will
lie at the binary orbital separation. The material spreads both
inwards and outwards from there. The steady state solutions calculated
in this work may only be appropriate inside of the initial binary
separation radius. Depending upon the spreading rate of the disk, the
dead zone may extend to the outer edge of the disk. This is especially
likely for high accretion rates, early in the disk evolution. The
spreading of the disk may be slowed because of the dead zone. We
discuss this further in Section~\ref{discussion}.

\subsection{Viscous heating}

\begin{figure}
\begin{center}
\includegraphics[width=8.4cm]{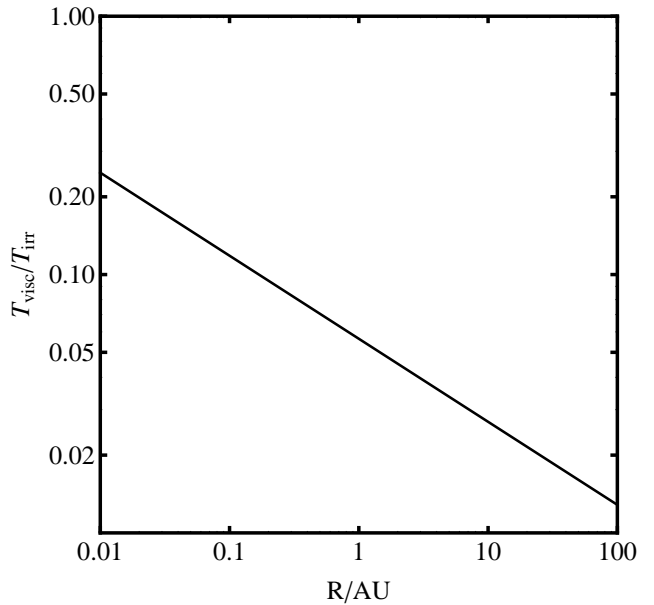}
\end{center}
\caption{Ratio of the temperature due to viscous heating to the
  irradiation temperature as a function of radius from the pulsar for
  a steady--state disk with accretion rates $\dot M=10^{-8}\,\rm
  M_\odot\, yr^{-1}$. }
\label{tv}
\end{figure}

In this work so far we have not included the effect of viscous heating
on the steady state disk model. We can calculate the temperature from
the viscous heating
\begin{equation}
\sigma T_{\rm visc}^4=\frac{9}{8}\nu\Sigma \Omega^2.
\end{equation}
With this, we calculate (for accretion rates below the Eddington
limit) the relative temperature compared with the irradiation temperature
\begin{align}
\frac{T_{\rm visc}}{T_{\rm irr}}=&\,\, 0.056
\left(\frac{\dot M}{10^{-8}\,\rm M_\odot\, yr^{-1}}\right)^{-\frac{1}{28}}
\left(\frac{R_{\rm p}}{10\,\rm km}\right)^{\frac{2}{7}}
\left(\frac{M_{\rm p}}{1.4\,\rm M_\odot }\right)^{\frac{3}{28}} \cr
&\,\,\times
\left( \frac{R}{1\,\rm AU}\right)^{-\frac{9}{28}}.
\end{align}
In Fig.~\ref{tv} we show this ratio as a function of radius for an
accretion rate of $10^{-8}\,\rm M_\odot\, yr^{-1}$. As seen from the
above equation, this is not very sensitive to the accretion
rate. Furthermore, the heating from viscous effects is much smaller
than the heating due to the irradiation from the pulsar. If we were to
include this heating into our steady state disk models, the
temperature of the disk would increase slightly. The models we
have shown here represent the maximum size of the dead zone in the
absence of viscous heating.

\section{Discussion}
\label{discussion}

\cite{Currie2007} considered the formation of pulsar planets in disks
with a dead zone for two scenarios, the supernova fallback disk and
the tidal disruption disk. They favor the supernova fallback disk for
planet formation because they suggest that it can form the compact
configuration of the observed planets.  However, recent numerical
simulations find that the formation of a fallback disk around a
neutron stars requires somewhat tuned parameters \citep{Perna2014}. As
we have shown in Section~\ref{22}, when the supernova fallback disk
does form, its lifetime is very short and would not allow for planet
formation. In the present work we have extended the parameter space of
dead zone properties. Most importantly, we have explored the
consequences of a broader range in the active layer surface
density. Since two of the planets in PSR B1257+12 are in a resonance,
this suggests that they formed farther out in the disk and migrated
inwards. Thus, we do not require that the dead zone would be in the
orbital location of the current locations of the planets. The planets
could form at an orbital radius of around $1\,\rm AU$ and then migrate
inwards to their observed locations. Formation farther out is also
possible, but this scenario should be investigated in a
time--dependent disk model. Migration through a dead zone is likely to
be much slower than migration through a fully viscous disk
\citep{Ward1997,Thommes2005,Matsumura2006} and so the planets can more
easily survive the migration process without being accreted on to the
pulsar in the presence of a dead zone. The inner edge of the dead zone
will move inwards in time as the disk mass and accretion rate drop
(see Fig.~\ref{rcrit}) and so the planets will likely remain in the
dead zone throughout the disk lifetime (if the dead zone exists).

\cite{Hansen2009pulsars} used the surface density profiles from the
layered disk models of \cite{Currie2007} to determine initial
conditions for the solid material in planetary embryos for a set of
N--body simulations to model the formation of planets around the
pulsar PSR B1257+12. They found that the layered disk model provided a
better fit to the final planetary systems because it resulted in the
planets forming from a narrow annulus \citep[this is also true in the
  solar system, see][]{Hansen2009b}.

Searches for debris disks around pulsars have revealed no infrared
counterparts \citep{Lohmer2004, Wang2014,Wang2014b}. However, debris
disks have been suggested to be around two magnetars 4U 0142+61 and 1E
2259+286 \citep{Wang2006,Kaplan2009}. Maybe the fact that planets
were not found yet orbiting white dwarfs
\citep[e.g.][]{Livio2005,Sandhaus2016, Farihi2016}, but that the
surface layers of some white dwarfs show (perhaps) pollution by
planets \citep[e.g.][]{Gansicke2012}, indicate that indeed there were
no dead zones, and either planets did not form, or else they migrated
all the way in through a viscous disk.

In this work we have concentrated on a steady--state disk
model. However, as we have discussed in Section~\ref{pd}, a disk model
that includes a dead zone is not in steady--state since material
builds up within the dead zone. Furthermore, the dead zone will slow
the rate of expansion of the disk \citep[e.g.][]{Hansen2002}. Material
only flows outwards in a layer of surface density $\Sigma_{\rm a}$
rather than the full surface density. For small active layer surface
densities, the disk will have a large extended region of low density
and a compact high density dead zone region. The steady--state disk
solutions described in this work are certainly adequate up to around
the initial separation of the binary, but outside of that,
time--dependent calculations will be required to determine the
evolution.  The non--monotonic surface density profile of the disk
will also affect the process of planet formation and may aid the
formation of a compact planetary system \citep{Hansen2009b}.  In a
future publication, we will investigate the time--dependent effects of
such a disk model and their potential implications for pulsar planet
formation.

\section{Conclusions}

We explain the small number of observed pulsar planets through a
combination of two low--probability events. First, the most likely
formation site is in a disk formed by the destruction of a companion
star. This also explains why all the pulsar planets that we have found
are around millisecond pulsars rather than young pulsars. A binary
that allows for such a scenario must have an extreme mass ratio and
has a very small chance of forming and surviving the supernova
explosion.  Second, planet formation is thought to require a dead
zone, a region of low turbulence, within the protoplanetary
disk. Because pulsars have a much stronger source of irradiation than
a young star, the additional heating can lead to sufficient ionization
that the dead zone does not form. The inner boundary of the dead zone
is pushed farther out, due to the irradiation, into a region of the
disk with lower surface density.  A dead zone only forms if the
surface density that is ionized by external sources is small,
$\Sigma_{\rm crit}\lesssim 10\,\rm g\, cm^{-2}$.

\section*{Acknowledgments} 

We thank Stephen Lepp for useful conversations. This research has made
use of the Exoplanet Orbit Database and the Exoplanet Data Explorer at
exoplanets.org.

\bibliographystyle{apj} \small


\label{lastpage}
\end{document}